\documentclass[conference]{IEEEtran}
\IEEEoverridecommandlockouts

\usepackage{cite}
\usepackage{amsmath,amssymb,amsfonts}
\usepackage{algorithmic}
\usepackage{graphicx}
\usepackage{textcomp}
\usepackage{xcolor}
\usepackage{booktabs}
\usepackage{tabularx}
\usepackage{multirow}
\def\BibTeX{{\rm B\kern-.05em{\sc i\kern-.025em b}\kern-.08em
    T\kern-.1667em\lower.7ex\hbox{E}\kern-.125emX}}

\usepackage{multirow}

\begin{document}

\title{Variational Autoencoder for \\ Personalized Pathological Speech Enhancement\\
\thanks{This work was supported by the Swiss National Science Foundation project 200021\_215187 on ``Pathological Speech Enhancement".}
}

\author{\IEEEauthorblockN{Mingchi Hou$^{1 ,2}$, Ina Kodrasi$^{1}$}
\IEEEauthorblockA{
\textit{$^1$Idiap Research Institute, Switzerland} \\
\textit{$^2$École Polytechnique Fédérale de Lausanne (EPFL), Switzerland}\\}

{\tt \{mingchi.hou,ina.kodrasi\}@idiap.ch}

}

\maketitle
\begin{abstract}

The generalizability of speech enhancement (SE) models across speaker conditions remains largely unexplored, despite its critical importance for broader applicability. This paper investigates the performance of the hybrid variational autoencoder (VAE)-non-negative matrix factorization (NMF) model for SE, focusing primarily on its generalizability to pathological speakers with Parkinson's disease. We show that VAE models trained on large neurotypical datasets perform poorly on pathological speech. While fine-tuning these pre-trained models with pathological speech improves performance, a performance gap remains between neurotypical and pathological speakers. To address this gap, we propose using personalized SE models derived from fine-tuning pre-trained models with only a few seconds of clean data from each speaker. Our results demonstrate that personalized models considerably enhance performance for all speakers, achieving comparable results for both neurotypical and pathological speakers.


\end{abstract}

\begin{IEEEkeywords}
speech enhancement, variational autoencoder, generalizability, Parkinson's disease, personalization
\end{IEEEkeywords}

\section{Introduction}
Speech communication is essential for conveying information, ideas, and emotions. However, noise from sources like traffic, machinery, or crowded environments impairs speech intelligibility and quality, posing challenges for many voice-based applications such as hearing aids and speech recognition systems~\cite{oshaughnessy_speech_2024}. 
To mitigate these challenges, speech enhancement (SE) approaches focusing on suppressing undesired interferences have become indispensable~\cite{SEbook-2013}.

In recent years, deep learning has revolutionized SE, leading to data-driven techniques that leverage large datasets and powerful learning algorithms to improve speech quality~\cite{RNN2017,CRN2018,CNNGRU_mask}. 
However, such approaches typically require pairs of clean and noisy data for training, and their performance heavily depends on the quantity and diversity of the training samples.
As a result, they often struggle to generalize in noisy environments that are not encountered during training~\cite{generalization_gap2023}. 
To achieve robustness in various noisy environments, generative models have recently become more prevalent~\cite{lu_cdiffuse_2022, richter2023sgmse, lemercier24spm, jukic2024sb,RVAE, STCN, NAVAE, DVAE, VAE_theory,jp2018}. These models can be broadly categorized into three categories, i.e., i) diffusion-based~\cite{lu_cdiffuse_2022, richter2023sgmse, lemercier24spm}, ii) Schrödinger bridge-based~\cite{jukic2024sb}, and iii) hybrid variational autoencoder (VAE) and non-negative matrix factorization (NMF)-based~\cite{RVAE, STCN, NAVAE, DVAE, VAE_theory,jp2018} models. Diffusion-based models operate by progressively adding noise to a clean speech signal in a forward process. 
In the reverse process, typically guided by a neural network, these models learn to reconstruct the clean signal from the noisy input. The more recent Schrödinger bridge-based generative model, in contrast, enables exact interpolation between the clean and noisy speech spectral components, moving beyond the conventional forward-backward noise process seen in diffusion models. Lastly, hybrid VAE-NMF-based models leverage a probabilistic latent space to model the clean speech prior through a VAE and an NMF to model the signal's structure dynamically, separating speech from noise in a more interpretable and structured way.
Although diffusion-based and Schr{\"o}dinger bridge-based models have generally shown better SE performance \cite{richter2023sgmse,jukic2024sb}, VAE-NMF-based models remain relevant and advantageous as they only require clean signals for training and typically have smaller model sizes. 


Despite the beneficial enhancement performance, a vast majority of SE research is done using English data and recordings from neurotypical speakers exhibiting no speech disorders. The generalizability to other languages has received seldom attention~\cite{close_effect_2023}.
More importantly, to the best of our knowledge, the performance of SE models has never been investigated for pathological speakers with speech disorders. 
Pathological speech, produced by individuals with neurological conditions such as Parkinson's disease (PD) or Amyotrophic Lateral Sclerosis, exhibits irregularities in speech characteristics~\cite{Darley_dysarthria}, which leads to reduced intelligibility and increased susceptibility to noise. 
Research has shown that there is a considerable difference in the statistical distribution of pathological speech compared to neurotypical speech~\cite{PD_ITG,kodrasi_spectro-temporal_2020}. Consequently, the performance of SE models trained using neurotypical speech recordings is expected to degrade when applied to pathological speech.
Given the widespread prevalence of neurological disorders, affecting more than 1 billion individuals~\cite{WHO_2006}, analyzing the performance of SE models for pathological speakers and developing SE solutions tailored to such speakers is crucial.

In this paper, we investigate the SE performance of the hybrid VAE-NMF model~\cite{jp2018, VAE_theory} across different languages and for neurotypical and pathological speakers. 
We analyze cross-language generalizability using English and Spanish datasets, showing that the hybrid VAE-NMF model is sensitive to domain shifts caused by language differences. Additionally, we compare the SE performance on both neurotypical and pathological speakers using VAE models trained on a large dataset of only neurotypical speech as well as VAE models trained on a considerably smaller dataset that includes both neurotypical and pathological speech. Unsurprisingly, we observe a performance gap between neurotypical and pathological speakers, with poorer SE performance for pathological speakers. Finally, we propose fine-tuning and personalization strategies, with personalization yielding not only the best overall performance for both neurotypical and pathological speakers, but also a comparable performance for both groups of speakers.

\section{VAE-NMF for Speech Enhancement}
\emph{Mixture model.} \enspace 
In the short-time Fourier transform (STFT) domain, the complex noisy mixture $y_{ft}$ at frequency bin index $f$ and time frame index $t$ is given by
\begin{equation}
    y_{ft} = \sqrt{g_t}s_{ft} + n_{ft},
    \label{equation:eq1}
\end{equation}
with $s_{ft}$ the clean speech, $n_{ft}$ the additive noise, and $g_t \in \mathbb{R}_+$ a frequency-independent time-varying gain introduced to provide robustness to the varying loudness level of different speech signals typically used for training~\cite{VAE_theory}. 
Given the noisy mixture and assuming that the speech and noise spectral coefficients are uncorrelated and follow a circularly symmetric Gaussian distribution, the minimum mean square error estimator of the clean speech coefficients is given by the Wiener filter~\cite{VAE_theory}
\begin{equation}
    \hat{s}_{ft} = \frac{\hat{g}_t\hat{\sigma}_{s,ft}^2}{\hat{g}_t\hat{\sigma}_{s,ft}^2 + \hat{\sigma}_{n,ft}^2} y_{ft},
    \label{equation:eq2}
\end{equation}
with $\hat{g}_t$, $\hat{\sigma}_{s,ft}^2$ and $\hat{\sigma}_{n,ft}^2$ the estimated gain, clean speech variance, and noise variance, respectively.
As briefly outlined in the following, the speech and noise variances can be estimated using a Monte Carlo expectation maximization (MCEM) algorithm combining a VAE which learns the prior distribution of clean speech and an untrained NMF-based noise model~\cite{VAE_theory}.

\emph{Clean speech prior.} \enspace 
The VAE consists of an encoder network $\mathbf{e}_{\phi}$, parametrized by $\phi$, and a decoder network $\mathbf{d}_{\theta}$, parameterized by $\theta$, operating on the squared magnitude of the STFT coefficients, i.e., the $F$-dimensional vector of speech power coefficients $|\mathbf{s}_t|^2$ at time frame index $t$, with $F$ being the total number of frequency bins.
The encoder maps $|\mathbf{s}_t|^2$ to the parameters of the distribution over the latent variable $\mathbf{z}_t \in \mathbb{R}^{D}$, where $D$ is the dimensionality of the latent space.
The decoder then reconstructs the clean speech power coefficients from the latent space. 
The likelihood of the clean speech coefficients given the latent variables is modeled as
\begin{equation}
p_\theta(\mathbf{s}_t\vert \mathbf{z}_t) = \mathcal{N}_{\mathbb{C}} (\mathbf{0}, {\rm diag}(\mathbf{d}_{\theta}(\mathbf{z}_t))), \; \; \text{with} \; \; \mathbf{z}_t \sim \mathcal{N} (\mathbf{0}, \mathbf{I}).
\end{equation}
The network parameters $\theta$ and $\phi$ are optimized by maximizing the evidence lower bound on the clean speech log-likelihood. 
For additional details, the interested reader is referred to~\cite{VAE_theory}.


\emph{Noise model.} \enspace 
NMF is used to model the noise variance as
\begin{equation}
    \sigma_{n,ft}^2 = \{\mathbf{W}\mathbf{H}\}_{ft},
    \label{equation:eq4}
\end{equation}
where $\mathbf{W} \in \mathbb{R}_+^{F \times K}$ and $\mathbf{H} \in \mathbb{R}_+^{K \times T}$ are non-negative matrices representing the temporal activations and noise basis, with $K$ being the NMF rank.

\emph{Noisy mixture model.} \enspace
Using the previously described clean speech and noise models, the distribution of the noisy mixture coefficients given the latent variable $\mathbf{z}_t$ follows 
\begin{equation}
    y_{ft}|\mathbf{z}_t \sim \mathcal{N}_{\mathbb{C}} (0,g_t \{ \mathbf{d}_{\theta}  (\mathbf{z}_t) \}_f + \{\mathbf{W}\mathbf{H}\}_{ft}),
    \label{equation:eq5}
\end{equation}
with the parameters $g_t$, $\mathbf{W}$, and $\mathbf{H}$ the unknown parameters to be estimated through MCEM~\cite{VAE_theory}.

\emph{SE with the hybrid VAE-NMF model.} \enspace
Given the noisy mixture signal, the gain $g_t$ and the speech and noise spectral variances are estimated as previously described. 
These estimated parameters are then used to compute the Wiener gain in~(\ref{equation:eq2}) and enhance the noisy signal.
Such a hybrid VAE-NMF approach to SE has been shown to achieve advantageous performance for neurotypical speech recordings~\cite{VAE_theory}.
Given the different distributions of neurotypical and pathological speech spectral coefficients~\cite{PD_ITG,kodrasi_spectro-temporal_2020}, it can be expected that the clean speech prior learned through a VAE trained on a neurotypical speech dataset does not generalize well to pathological speech.




\section{Personalized \\ Pathological Speech Enhancement}
\label{sec: personal}
It is already known that SE models exhibit highly variable performance due to mismatches in speaker characteristics between training and testing sets~\cite{Indiana_interspeech2021,Illinois_interspeech2014}, although this has not been explored for pathological speakers or in the context of the hybrid VAE-NMF model. 
Personalized SE models, which optimize performance for a single speaker within a specific acoustic environment, have been shown to outperform general-purpose SE models for that particular speaker and environment~\cite{TASLP2017,Indiana_WASPAA2021}. 
Several approaches have been proposed to adapt general SE models to test speakers, including incorporating speaker embeddings into models like Deep Complex Convolution Recurrent Network and Deep Convolution Attention U-Net~\cite{MS_ICASSP2022}, training speaker encoder networks~\cite{Tencent_ICASSP2022}, and using internal embeddings to capture speaker profiles~\cite{MS_Interspeech2024}. Another approach involves fine-tuning pre-trained SE models on a pseudo-SE task, using only noisy signals from the test speaker of interest~\cite{Indiana_interspeech2021}.

In exploring strategies to enhance the hybrid VAE-NMF model's performance on pathological speech, we consider two approaches. First, we investigate the impact of including pathological speech data during VAE training. Instead of random initialization, we fine-tune a pre-trained VAE model (originally trained on a large dataset of neurotypical speakers) on a smaller dataset containing both neurotypical and pathological speech.
As shown in Section~\ref{sec: exp_ft_p}, while fine-tuning the VAE on a pathological dataset improves the SE performance of the hybrid VAE-NMF model for pathological speakers, a performance gap between neurotypical and pathological speaker remains. 
We argue that the characteristics of pathological speech vary widely between speakers, and including pathological speech in the VAE training set does not necessarily improve generalization to unseen pathological test speakers.
To further improve performance, our second approach involves using a personalized SE model for each speaker.
Given the challenges of obtaining robust embeddings for pathological speakers, we follow a personalization approach similar to~\cite{Indiana_interspeech2021}, where signals from the test speaker are used to fine-tune the pre-trained SE model.
Differently from~\cite{Indiana_interspeech2021} where fine-tuning is done using noisy signals, we assume availability of clean signals from the test speakers (since the VAE is trained on clean speech).

\section{Experimental Settings}\label{sec:experimental_settings}
\subsection{Datasets}\label{subsec:datasets}

Clean speech signals are used for training and validation of VAE models, while noisy mixtures are used for testing.

{\emph{Clean signals.}} \enspace To assess cross-language generalizability, we use the English Wall Street Journal (WSJ0)~\cite{garofolo_john_s_csr-i_2007_wsj0} and the Spanish Crowdsourced Latin American Spanish Corpora (CROWD)~\cite{guevara-rukoz_crowdsourcing_2020} datasets, which contain recordings from neurotypical speakers. The characteristics of these datasets, including the number of speakers and total duration of the training, validation, and test sets, are presented in Table~\ref{tab:dataset_info}. As shown, the datasets are closely matched in terms of speaker count and total duration, ensuring a systematic evaluation of cross-language generalizability. 
To assess cross-speaker generalizability, fine-tune models, and develop personalized models, we use the Spanish PC-GITA dataset~\cite{orozco-arroyave_new_2014_pcgita}, comprising recordings from $50$ neurotypical and $50$ PD speakers. PC-GITA includes recordings of $10$ sentences, one text passage, and a monologue per speaker, and has a total duration of $2.8$ hours. Due to this dataset's limited size (which is a common characteristic of pathological speech datasets) compared to those used in SE literature (cf. Table~\ref{tab:dataset_info}), we use a $10$-fold speaker-independent cross-validation evaluation framework, partitioning the data within each fold into $80$\% training, $10$\% validation, and $10$\% testing. Personalized models are trained, validated, and tested per speaker independently (cf. Section~\ref{sec: set_ft_p}).


\begin{table}[t]
\centering
\caption{WSJ0 (English) and CROWD (Spanish) dataset characteristics.}
\label{tab:dataset_info}
\begin{tabularx}{0.5\textwidth}{Xll}
\hline
& WSJ0~\cite{garofolo_john_s_csr-i_2007_wsj0} & CROWD~\cite{guevara-rukoz_crowdsourcing_2020} \\ \hline
Training set & $101$ speakers, $24.9$ hours & $101$ speakers, $23.0 $ hours \\ \hline
Validation set & $10$ speakers, $2.2$ hours & $10$ speakers, $2.2 $ hours \\ \hline
Test set & $8$ speakers, $1.5$ hours & $8$ speakers, $1.5 $ hours  \\ \hline
\end{tabularx}
\end{table}

\emph{Noisy mixtures.} \enspace 
Noisy mixtures required for testing the VAE-NMF model are generated using the previously described clean signals and noise recordings from the QUT dataset~\cite{qut}. The noise recordings are randomly selected from the set \{``cafe", ``car", ``home", ``street"\}, with the signal-to-noise ratio (SNR) randomly selected from the set \{-5, 0, 5\} dB~\cite{STCN}.

\subsection{Training}\label{subsec:training_configurations}
All signals are resampled to $16$ kHz if necessary and are transformed to the STFT domain using a $64$ ms Hann window with a hop size of $16$ ms, resulting in $513$ unique frequency bins.
As in~\cite{RVAE,NAVAE}, the VAE latent dimension is $D=16$ and the NMF rank is $K = 8$.
The settings of the MCEM follow~\cite{VAE_theory}.
Additionally, we use a batch size of $128$, the Adam optimizer with standard settings, and an initial learning rate of $10^{-4}$.
The learning rate scheduler halves the learning rate if the validation loss plateaus, with a patience of $10$ epochs.
The maximum number of epochs is set to $500$, with early stopping if the validation loss does not decrease for $20$ consecutive epochs.

\begin{table*}
\footnotesize
    \centering
    \caption{Within-database and cross-database performance of the VAE-NMF SE models.  VAE (WSJ0) denotes models trained on the English WSJ0 dataset, whereas VAE (CROWD) denotes models trained on the Spanish CROWD dataset. Models are tested on English (WSJ0-QUT) and Spanish (CROWD-QUT) test sets. Values in bold indicate the test set for which the highest performance improvement is obtained for each model and each measure.}
    \label{tbl: lang}
    \begin{tabularx}{\textwidth}{X|rrr|rrr}
    \toprule
     Model & \multicolumn{3}{c|}{English (WSJ0-QUT) test set} & \multicolumn{3}{c}{Spanish (CROWD-QUT) test set} \\
      & $\Delta$PESQ & $\Delta$fwSSNR & $\Delta$SI-SDR 
     &  $\Delta$PESQ & $\Delta$fwSSNR & $\Delta$SI-SDR \\
    \midrule
     VAE (WSJ0) - English train set  & $ \bf 0.21 \pm 0.02 $ & $ \bf 0.87 \pm 0.13 $ & $ \bf 5.77 \pm 0.28 $ & $ 0.13 \pm 0.01 $ & $ -0.91 \pm 0.10 $ & $ 4.53 \pm 0.25 $ \\
     VAE (CROWD) - Spanish train set & $ 0.13 \pm 0.01 $ & $ 0.69 \pm 0.13 $ & $ 4.65 \pm 0.26 $ & $ \bf 0.19 \pm 0.01 $ & $ \bf 0.78 \pm 0.08 $ & $ \bf 6.25 \pm 0.22 $ \\
    \bottomrule
    \end{tabularx}
\end{table*}
\begin{table}[t!]
\footnotesize
    \centering
    \caption{Performance of the VAE (CROWD) model on the PC-GITA-QUT test set of neurotypical and pathological speakers.  Values in bold indicate the speaker group for which the highest performance improvement is obtained for each measure.}
    \label{tbl: patho}
    \begin{tabularx}{0.5\textwidth}{X|rrr}
    \toprule
    Test Speakers & $\Delta$PESQ & $\Delta$fwSSNR & $\Delta$SI-SDR \\
    \midrule
    Neurotypical & $\bf0.10 \pm 0.01$ & $\bf2.22 \pm 0.14$ & $\bf4.72 \pm 0.26$ \\
    Pathological & $0.05 \pm 0.01$ & $1.51 \pm 0.17$ & $3.76 \pm 0.30$ \\
    \bottomrule
    \end{tabularx}
\end{table}

\begin{table}[t!]
\footnotesize
    \centering
    \caption{Performance of different models on the PC-GITA-QUT test set of neurotypical and pathological speakers. Values in bold indicate the speaker group for which the highest performance improvement is obtained for each measure.}
    \label{tbl: personal}
    \addtolength{\tabcolsep}{-0.05em}
    \begin{tabularx}{0.5\textwidth}{X|r|rrr}
    \toprule
    Model & Test Speakers & $\Delta$PESQ & $\Delta$fwSSNR & $\Delta$SI-SDR \\
    \midrule
    \multirow{2}{4em}{$\text{M}_\text{S}$} & Neurotypical  & $0.13 \pm 0.01$ & $1.96 \pm 0.19$ & $4.48 \pm 0.35$ \\
    & Pathological  & $0.07 \pm 0.01$ & $1.08 \pm 0.20$ & $3.22 \pm 0.37$ \\
    \midrule
    \multirow{2}{4em}{$\text{M}_\text{F}$} & Neurotypical & $0.13 \pm 0.01$ & $2.53 \pm 0.16$ & $5.02 \pm 0.29$ \\
    & Pathological & $0.08 \pm 0.01$ & $1.79 \pm 0.17$ & $4.10 \pm 0.32$ \\
    \midrule
    \multirow{2}{4em}{$\text{M}_\text{P}$} & Neurotypical & $\bf 0.19 \pm 0.02$ & $\bf 2.74 \pm 0.16$ & $\bf 7.87 \pm 0.34$ \\
    & Pathological & $0.18 \pm 0.02$ & $2.40 \pm 0.16$ & $7.75 \pm 0.37$ \\
    \bottomrule
    \end{tabularx}
\end{table}

\subsection{Fine-tuning and personalization}
\label{sec: set_ft_p}
Given the small size of the PC-GITA dataset, training a VAE model from scratch (i.e., with random weight initialization) on this dataset is expected to yield suboptimal results. To improve performance, we explore the possibility of leveraging the pre-trained VAE model from the CROWD dataset and fine-tuning it on the PC-GITA dataset. After initializing the model weights, fine-tuning is done following the same training procedure outlined in Section~\ref{subsec:training_configurations}.

As outlined in Section~\ref{sec: personal}, we also train personalized models for each speaker in the PC-GITA dataset. To this end, we use a subset of a speaker's recordings as training/validation data, while the remaining recordings are used as testing data. To ensure that the conclusions we draw from these models are not influenced by the specific subset of data used for training/validation/testing, we report the average performance of two personalized models: one using the monologue recording for training/validation and the sentences and read text recordings for testing, and another using the sentences and read text recordings for training/validation and the monologue recording for testing.
It should be noted that these subsets of data have a similar duration, i.e., the average duration of the monologue recording across all speakers is $47.1$ s whereas the average duration of the sentences and read text recording across all speakers is $55.3$ s.
Personalized models are initialized with the weights of the pre-trained VAE
model from the CROWD dataset.
The training procedure is the same as outlined in Section~\ref{subsec:training_configurations}.

\subsection{Evaluation}\label{subsec:evaluation_metrics}
Performance is evaluated using the wideband perceptual evaluation of speech quality (PESQ) measure~\cite{pesq_2001}, the frequency-weighted segmental SNR (fwSSNR)~\cite{fwSSNR}, and the scale-invariant signal-to-distortion ratio (SI-SDR)~\cite{roux_sdr_2019}. 
The clean speech signal is used as a reference signal for computing these measures.
The improvement in these instrumental measures, i.e., $\Delta$PESQ, $\Delta$fwSSNR, and $\Delta$SI-SDR, is computed as the difference between the PESQ, fwSSNR, and SI-SDR values of the enhanced signal and the noisy microphone signal.

\section{Results and Discussion}
\subsection{Cross-lingual generalization} \label{sec: cross}
Table~\ref{tbl: lang} presents the performance of the hybrid VAE-NMF model with the VAE trained on the English WSJ0 and Spanish CROWD datasets, evaluated under both within-database and cross-lingual testing conditions. It should be noted that the noise types and SNRs remain consistent across all evaluations. However, since the VAE is trained on the different considered clean speech datasets, domain shifts are introduced due to different languages (and potentially different recording setups). As expected, the VAE achieves a better performance in within-database testing (i.e., when the training and testing datasets are in the same language). In contrast, performance degrades in cross-lingual settings due to the domain shift introduced by the different language.

\subsection{Cross-speaker generalization} 
In this section, we assess the performance of the VAE-NMF model on both neurotypical and pathological speakers from the PC-GITA database. Since PC-GITA consists of Spanish recordings, our analysis focuses on the VAE (CROWD) model, i.e., the model trained on the Spanish CROWD dataset in Section~\ref{sec: cross}. The results, presented in Table~\ref{tbl: patho}, reveal that the model performs consistently better on neurotypical speakers than on pathological speakers across all evaluation metrics. This degradation is expected, as VAE (CROWD) is trained exclusively on neurotypical speech, with pathological speech characteristics exhibiting substantial deviations from it~\cite{kodrasi_spectro-temporal_2020}.

\subsection{Fine-tuning and personalization}
\label{sec: exp_ft_p}
In this section, we explore various strategies to improve the SE performance of the VAE-NMF model for pathological speakers. Specifically, we investigate the performance of the following models:
\begin{itemize}
\item {\emph{Training a VAE from scratch on PC-GITA (model $\text{M}_\text{S}$).}} \enspace We examine the feasibility of training a VAE model from scratch on a small, clean speech dataset, such as PC-GITA, which includes both neurotypical and pathological speakers. This contrasts with the conventional practice of training VAEs on much larger corpora, such as WSJ0 or CROWD, which consist of only neurotypical speech.
\item {\emph{Fine-tuning a pre-trained VAE (model $\text{M}_\text{F}$).}} \enspace We explore the potential of fine-tuning a VAE model that has been pre-trained on the larger CROWD dataset, using the PC-GITA dataset, as described in Section~\ref{sec: set_ft_p}.
\item {\emph{Personalized VAE model (model $\text{M}_\text{P}$).}} \enspace We assess the performance of personalized VAE models, where the model is trained for each individual speaker as outlined in Section~\ref{sec: set_ft_p}.
\end{itemize}
The SE performance of the different models for both neurotypical and pathological speakers is summarized in Table~\ref{tbl: personal}.
Results show that training the VAE solely on the PC-GITA dataset leads to a poorer performance in terms of fwSSNR, and SI-SDR for both neurotypical and pathological speakers, compared to training the VAE on the larger CROWD dataset (cf. Table~\ref{tbl: patho}). This highlights the importance of using a sufficiently large and diverse dataset when training VAEs.
Rather than training the VAE from scratch on PC-GITA, Table~\ref{tbl: personal} demonstrates that fine-tuning the VAE model pre-trained on CROWD using the PC-GITA dataset leads to improvements across all metrics for both groups of speakers. However, a persistent performance gap remains between neurotypical and pathological speakers. We hypothesize that the high variability in pathological speech across individuals restricts the model’s ability to generalize effectively to new pathological speakers, even when pathological speech is included in the training data.
Finally, Table~\ref{tbl: personal} highlights the value of using personalized models. Taking advantage of the knowledge embedded in the pre-trained VAE (CROWD), incorporating only a few seconds ($\approx 50$ s) of clean speech from the test speaker can boost the SE performance. While this personalization strategy proves effective for both groups of speakers, the improvement in SE performance is particularly notable for pathological speakers.

\section{Conclusion}

This paper evaluates the generalizability of the hybrid VAE-NMF model for SE across languages (i.e., English and Spanish) and speaker conditions (i.e., neurotypical speakers and pathological speakers with PD). The results show that, as expected, the hybrid VAE-NMF performs best when trained and tested within the same database. However, cross-lingual and cross-speaker testing leads to performance degradation, with a larger drop observed in the latter case. To improve performance for pathological speech, we have proposed fine-tuning pre-trained models by incorporating pathological data into the training set, as well as training personalized models for each speaker using only a few seconds of clean data. Results have shown that personalized SE models significantly improve performance for all speakers, achieving comparable results for both neurotypical and pathological speakers.






\bibliographystyle{IEEEtran}
\bibliography{mybib}

\end{document}